\documentclass[prc,byrevtex,showpacs,twocolumn,amsmath,amssymb,superscriptaddress]{revtex4}

\usepackage{graphicx}
\usepackage{dcolumn}
\usepackage{bm}
\usepackage{multirow}
\usepackage{longtable}

\newcolumntype{.}{D{.}{.}{2}}

\begin{document}

\newcommand{\D}{\displaystyle}

\title{Gold fragmentation induced by stopped antiprotons}

\author{P. Lubi\'nski}
\email{piotr@camk.edu.pl}
\affiliation{Heavy Ion Laboratory, Warsaw University, Pasteura 5a, 02-093 Warsaw, Poland}
\affiliation{Nicolaus Copernicus Astronomical Center, Bartycka 18, 00-716 Warsaw, Poland}
\author{A. Grochulska}
\affiliation{Heavy Ion Laboratory, Warsaw University, Pasteura 5a, 02-093 Warsaw, Poland}
\author{T. von Egidy}
\affiliation{Physik-Department, Technische Universit\"at M\"unchen, D-8046 Garching, Germany}
\author{K. Gulda}
\affiliation{Institute of Experimental Physics, Warsaw University, Ho\.za 69, 00-681 Warsaw, Poland}
\author{F.J. Hartmann} 
\affiliation{Physik-Department, Technische Universit\"at M\"unchen, D-8046 Garching, Germany}
\author{J.~Jastrz\c{e}bski}
\affiliation{Heavy Ion Laboratory, Warsaw University, Pasteura 5a, 02-093 Warsaw, Poland}
\author{W. Kurcewicz}
\affiliation{Institute of Experimental Physics, Warsaw University, Ho\.za 69, 00-681 Warsaw, Poland}
\author{L. Pienkowski}
\affiliation{Heavy Ion Laboratory, Warsaw University, Pasteura 5a, 02-093 Warsaw, Poland}
\author{A. Stolarz}
\affiliation{Heavy Ion Laboratory, Warsaw University, Pasteura 5a, 02-093 Warsaw, Poland}
\author{A. Trzci\'nska}
\affiliation{Heavy Ion Laboratory, Warsaw University, Pasteura 5a, 02-093 Warsaw, Poland}

\date{\today}

\begin{abstract}

A natural gold target was irradiated with the antiproton beam from the Low Energy Antiproton 
Ring at CERN. Antiprotons of 200 MeV/c momentum were stopped in a thick target, products of 
their annihilations on Au nuclei were detected using the off-line $\gamma$-ray spectroscopy 
method. In total, yields for 114 residual nuclei were determined, providing a data set to 
deduce the complete mass and charge distribution of all products with $A$ $\gtrsim$ 20 from 
a fitting procedure. The contribution of evaporation and fission decay modes to the total 
reaction cross section as well as the mean mass loss were estimated. The fission probability for Au 
absorbing antiprotons at rest was determined to be equal to ($3.8\pm 0.5$) \%, in good 
agreement with an estimation derived using other techniques. The mass-charge yield distribution was 
compared with the results obtained for proton and pion induced gold fragmentation. On the average, 
the energy released in $\rm \bar{p}$ annihilation is similar to that introduced  by $\approx$1 
GeV protons. However, compared to proton bombardment products, the yield distribution of antiproton 
absorption residues in the $N$-$Z$ plane is clearly distinct. The data for antiprotons exhibit also 
a substantial influence of odd-even and shell effects.

\end{abstract}

\pacs{25.43.+t, 25.85.Ge}

\maketitle

\section{\label{introduction}Introduction}

The large energy of almost 2 GeV released in nucleon-antinucleon annihilation had awoken hopes
to observe some unique nuclear reactions induced in this way. Especially energetic antiprotons were
presumed to reach the deep interior of the nucleus. Exotic processes, like phase transitions to quark-gluon 
plasma and explosive decay of that hot system had been expected to occur \cite{Rafelski80,Clover82},
but were not verified in experiments performed shortly after the commissioning of the Low Energy Antiproton 
Ring at CERN
\cite{Balestra86,McGaughey86}. Nevertheless, the character of the reactions starting with 
antiproton absorption in nuclei is quite unique in comparison with reactions induced with 
protons or heavy ions. Whereas the excitation energy carried in by post-annihilation mesons
is quite large, the linear and angular momentum transfer as well as the matter compression are reduced,
particularly in the stopped antiproton case which is preceded by the exotic atom phase. Hence, one may
investigate a clear thermal reaction aspects with suppressed collective motion complications. 

Such phenomena were intensively studied during the LEAR era for stopped and for energetic
antiprotons. The spectra for neutrons and light charged particles 
\cite{Markiel88,Hofmann90,Machner92,Polster95,Goldenbaum96,Lott01}, mass yield distributions 
\cite{Moser86,Moser89,Egidy90,Jastrzebski93,Szmalc92,Gulda93} and characteristics of the fission fragments 
\cite{Hofmann94,Kim96,Jahnke99} were measured for a wide range of targets.
The mean excitation energy derived from these studies, $\approx$ 150 MeV and $\approx$ 300 MeV for 
heavy nuclei absorbing stopped and 1.2 GeV antiprotons, respectively, compares well with the average
values obtained for protons which have about 1 GeV larger energies, i.e. approximately the nucleon
rest mass.

Yield distributions of residual nuclei were studied as soon as more intense $\rm \overline{p}$ beams 
were provided by LEAR. Many targets were irradiated with antiprotons, mainly 
at rest energy, but only few of them were examined in detail: $^{\rm nat}$Cu \cite{Jastrzebski93}, 
$^{92,95}$Mo \cite{Moser86}, $^{98}$Mo \cite{Moser89}, $^{\rm nat}$Ag \cite{Szmalc92}, $^{\rm nat}$Ba 
\cite{Egidy90}, 
$^{165}$Ho \cite{Moser89} and $^{181}$Ta \cite{Gulda93}. Average quantities such as the mean mass 
removed from the 
target as well as individual yield features, e.g. isomeric ratios, were investigated. Clear odd-even 
effects in the mass and charge yield distribution were observed. Theoretical calculations, based on 
intranuclear cascade + evaporation models, were able to reproduce 
only the gross features and failed to predict yield dependence on the detailed $N$ and $Z$ \cite{Egidy90}. 
The odd-even phenomenon, although postulated to be present and used to model the yield distribution for
years \cite{Rudstam66,Silberberg73} and sometimes reaching very large values \cite{Tipnis98},
still seems to be almost completely unexplored theoretically in a more quantitative way.

One of the most distinct features observed in reactions with stopped antiprotons is the large probability 
($\approx$ 0.1) of very small energy transfer, when the target nucleus looses only one nucleon in 
annihilation and is left excited below the nucleon separation energy. Nuclear spectroscopy studies of the 
relative yield of both types of these residual ($A_{T}-1$) nuclei (a neutron or a proton lost in such soft
antiproton absorptions) were used to establish a new and powerful method of probing the nuclear periphery 
composition \cite{Jastrzebski93a,Lubinski94,Lubinski98,Schmidt99}.

The irradiation of the heavy, gold target gave us a chance to study the competition between evaporation 
and fission 
induced by antiproton absorption. The yield distribution of heavy residual nuclei complements the results 
obtained from on-line measurements of neutron and charged particle spectra \cite{Polster95}. Since gold 
is a commonly used target, there was also the opportunity to compare these data with a rich set of 
information 
gathered from irradiations with energetic protons or pions. In particular, the yield distribution after
the reaction of 1 GeV protons with Au has been extensively studied in older and recent $\gamma$-ray 
spectroscopy measurements \cite{Kaufman80,Michel97} and also with a new method using the mass-charge 
spectrometry for inverse kinematic 
reactions \cite{Rejmund01,Benlliure01}. Our preliminary data of the reaction $\overline{p}$ + Au at rest 
were already published \cite{Jastrzebski95}, this work presents the complete results obtained after 
fitting the $Y(A,Z)$ yield distribution.

The article is organized as follows. In Section \ \ref{sec:experiment} we briefly present some details 
of the experiment, in Sec.\ \ref{sec:analysis} data analysis is described together with the yield 
fitting method.
Experimental results are presented and initially discussed within Sec.\ \ref{sec:results}. 
In Sec.\ \ref{sec:discussion} we compare the results of this experiment with those obtained with
other projectiles impinging on Au and with data coming from studies of antiproton absorption in various 
targets. Finally, our main conclusions are presented in Sec.\ \ref{sec:conclusions}.

\section{EXPERIMENT}
\label{sec:experiment}

A thick target of natural gold was irradiated with the antiproton beam of 200 MeV/c momentum
from LEAR facility. The target of the total thickness of 549 mg/cm$\rm ^{2}$ was composed
of ten foils of 80, 30, 30, 30, 2, 30, 30, 37, 80 and 200 mg/cm$\rm ^{2}$, starting from the
beam side. The initial energy of the antiprotons, equal to about 21 MeV, was reduced in 
the scintillation counter (from Pilot B) and in some additional moderators (mylar,silicon) to about 6.5 MeV
at the target surface. Such an arrangement assured that the majority of antiprotons was stopped 
in few central Au foils. The very central and extremely thin foil of 2 mg/cm$\rm ^{2}$ was applied 
to monitor the X-ray activity, while the last and thickest one (200 mg/cm$\rm ^{2}$) was used 
to check the secondary reactions level.

Two scintillation counters, S1 and S2, were used to control beam intensity and transverse
dispersion. The first anticounter S1 had a hole of 10 mm diameter and the active area of counter S2 
was a disc of the same diameter. Consequently the signal $\rm \overline{S1}S2$ indicated particles 
going towards the target. The irradiation lasted 15 minutes with the total number of antiprotons equal to 
$(9.25\pm 0.35)\rm \times 10^{8}$ ($\rm \overline{S1}S2$ number).

Monitoring of the target activity started 13 min after the irradiation and continued at CERN during
one week; afterwards the spectra were collected in Warsaw, the last one
was taken more than a year after target activation. Two HPGe detectors were used at CERN, a $\gamma$-ray 
counter of 15\% relative efficiency for all foils and an X-ray counter for the thinest one. 
In Warsaw two more efficient $\gamma$-ray detectors were applied, of 20\% and of 60\% relative 
efficiency, and a third X-ray detector for the thin foil.

All collected spectra were analyzed with the program ACTIV \cite{Zlokazov82,Karczmarczyk90}.
Gamma-ray lines were identified by their energies, half-lives and intensity ratios. The decay 
data were taken from the eighth edition of the Table of Isotopes \cite{Firestone96}.

Experimental yields for all detected residual nuclei, normalized to 1000 $\overline{p}$ stopped in the 
target, are listed in Table \ref{yields}. The independent yields represent the total number of nuclei, 
summed over all isomers. Cumulative yields include also the yields of all $\beta$-decay precursors of 
a given isotope. Besides that, there are presented partial yields for some 
isomers not representing the whole production for a given (A,Z) pair as well as some production limits for 
Hg nuclei. Mercury may be produced from gold after $\rm \overline{p}$ absorption in a charge exchange 
reactions, when one of the annihilation $\pi ^{+},\pi ^{0}$ pions exchanges charge with a target 
neutron. Such a phenomenon was observed for some targets irradiated with stopped antiprotons 
\cite{Moser89,Egidy90,Szmalc92,Gulda93}, where nuclei of target charge plus one were produced at level 
ranging from 0.5 to 5 per 1000 $\rm \overline{p}$. On the other hand, for some other targets, studied in the 
neutron halo project, rather low upper limits (0.5-2\%) were given \cite{Lubinski97}. Our data 
obtained for Au, except for the $^{195}$Hg isomers with low $\gamma$ intensity, indicate that such an effect
should happen very rarely.

Initially, the distribution of the activity induced in individual target foils was estimated several hours after
the irradiation with the use of the measurement of the $\rm ^{186}Ir(g)$ yield in each foil. Five inner foils 
(30, 30, 2, 30, 30 mg/cm$\rm ^{2}$) gathered about 90\% of the total activity and only these foils were
monitored later. This reduced the $\gamma$-ray self-absorption effect. The distribution of the
target activity  
was determined more precisely afterwards on the basis of yields obtained for six evaporation residues:
$^{186}$Ir, $^{184}$Ir, $^{183}$Os, $^{181}$Re, $^{157}$Dy and $^{152}$Dy. On average, five inner foils
stopped ($89.7\pm 2.1$)\% of the whole number of antiprotons. These data, together with the results
for $^{196}$Au and $^{192}$Au, were used also to estimate the yield introduced by secondary 
reactions with particles (mainly pions and neutrons) produced after antiproton annihilation on target nuclei.
It was done by comparing the number of given nuclei produced per foil thickness unit, averaged
for three inner foils, with the similar result obtained for the last, thickest foil. The upper limit 
for the secondary reactions leading to $^{196}$Au is equal to 3\%, the limit for $^{192}$Au is about 2.6\% 
and for the rest of quoted isotopes it does not exceed 2\%, i.e. in all cases it is below the yield 
uncertainties. The negligible influence of the secondary reactions on our results is additionally
confirmed by very small upper limit given in Table \ref{yields} for $^{198}$Au, a (n,$\gamma$) reaction
product.

\begingroup
\setlength{\tabcolsep}{3pt}
\begin{longtable*}{lrd@{${}\pm{}$}..cclrd@{${}\pm{}$}..c}
\caption{Experimental and fitted yields of residual nuclei from gold fragmentation induced with
stopped antiprotons. Yield type: I - independent, C - cumulative, in brackets - total yield fitted
for isotopes for which only one isomeric state was observed.\label{yields}}\\
\colrule \colrule
\multicolumn{1}{c}{Nuclide} & \multicolumn{1}{r}{T$_{1/2}$} & \multicolumn{2}{c}{Experiment} &
\multicolumn{1}{c}{Fit} & \multicolumn{1}{c}{Type} & &
\multicolumn{1}{c}{Nuclide} & \multicolumn{1}{c}{Half-life} & \multicolumn{2}{c}{Experiment} &
\multicolumn{1}{c}{Fit} & \multicolumn{1}{c}{Type} \\
\multicolumn{2}{c}{} & \multicolumn{3}{c}{[N/1000$\rm \bar{p}$]} & \multicolumn{1}{c}{} & &
\multicolumn{2}{c}{} & \multicolumn{3}{c}{[N/1000$\rm \bar{p}$]} & \multicolumn{1}{c}{} \\
\colrule
\endfirsthead
\caption{- continued.}\\
\colrule
\multicolumn{1}{c}{Nuclide} & \multicolumn{1}{c}{Half-life} & \multicolumn{2}{c}{Experiment} &
\multicolumn{1}{c}{Fit} & \multicolumn{1}{c}{Type} & &
\multicolumn{1}{c}{Nuclide} & \multicolumn{1}{c}{Half-life} & \multicolumn{2}{c}{Experiment} &
\multicolumn{1}{c}{Fit} & \multicolumn{1}{c}{Type} \\
\multicolumn{2}{c}{} & \multicolumn{3}{c}{[N/1000$\rm \bar{p}$]} & \multicolumn{1}{c}{} & &
\multicolumn{2}{c}{} & \multicolumn{3}{c}{[N/1000$\rm \bar{p}$]} & \multicolumn{1}{c}{} \\
\colrule
\endhead
$^{198}$Au$^{\rm g}$  &   2.7 d &  \multicolumn{2}{d}{< 0.6} & \multicolumn{1}{c}{$\text{---}$} & I & &
$^{165}$Tm        &  30.1 h & 14.4  &  0.8  & 14.2                                      & C \\
$^{196}$Au$^{\rm m2}$ &   9.7 h &  1.52 &  0.18 & \multicolumn{1}{c}{$\text{---}$}              & I & &
$^{163}$Tm        &  1.81 h & 10.1  &  0.6  & 10.1                                      & C \\
$^{196}$Au        &   6.2 d & 75.0  &  3.4  & \multicolumn{1}{c}{$\text{---}$}                  & I & &
$^{161}$Tm        &  33.0 m &  8.6  &  1.1  &  8.6                                      & C \\
$^{195}$Hg$^{\rm m}$  &  41.6 h & \multicolumn{2}{d}{< 3.1} & \multicolumn{1}{c}{$\text{---}$}  & I & &
$^{161}$Er        &   3.2 h &  2.69 &  0.67 &  2.73                                     & I \\
$^{195}$Hg$^{\rm g}$  &   9.9 h & \multicolumn{2}{d}{< 11.7} & \multicolumn{1}{c}{$\text{---}$} & I & &
$^{160}$Er        &  26.6 h &  9.8  &  0.8  &  9.9                                      & C \\
$^{195}$Au        &   186 d & 38.7  &  3.1  & \multicolumn{1}{c}{$\text{---}$}                  & I & &
$^{159}$Er        &  36.0 m &  8.2  &  0.8  &  8.2                                      & C \\
$^{194}$Au        &  38.0 h & 11.8  &  0.6  & \multicolumn{1}{c}{$\text{---}$ }                 & I & &
$^{157}$Dy        &   8.2 h &  8.0  &  0.4  &  7.9                                      & C \\
$^{193}$Hg$^{\rm m}$  &  11.8 h & \multicolumn{2}{d}{< 0.5} & \multicolumn{1}{c}{$\text{---}$}  & I & &
$^{155}$Dy        &   9.9 h &  6.2  &  0.3  &  6.2                                      & C \\
$^{193}$Hg$^{\rm g}$  &   3.8 h & \multicolumn{2}{d}{< 2.8} & \multicolumn{1}{c}{$\text{---}$}  & I & &
$^{155}$Tb        &   5.3 d &  0.27 &  0.09 &  0.30                                     & I \\
$^{193}$Au        &  17.7 h &  9.9  &  2.2  & \multicolumn{1}{c}{$\text{---}$}                  & I & &
$^{154}$Ho$^{\rm g}$  &  11.8 m &  2.93 &  0.20 & (3.16)                                    & C \\
$^{192}$Hg        &   4.9 h & \multicolumn{2}{d}{< 1.3} & \multicolumn{1}{c}{$\text{---}$}      & I & &
$^{153}$Tb        &   2.3 d &  4.77 &  0.24 &  4.78                                     & C \\ 
$^{192}$Au        &   4.9 h &  8.5  &  0.6  &  8.37                                     & I & &
$^{153}$Gd        &   242 d &  0.13 &  0.04 &  0.12                                     & I \\
$^{192}$Ir$^{\rm g}$  &  73.8 d &  1.61 &  0.07 &  2.6                                      & I & &
$^{152}$Dy        &   2.4 h &  3.60 &  0.11 &  3.60                                     & C \\
$^{191}$Hg$^{\rm m}$  &  51.0 m & \multicolumn{2}{d}{< 1.7} & \multicolumn{1}{c}{$\text{---}$}  & I & &
$^{152}$Tb        &  17.5 h &  0.67 &  0.13 &  0.67                                     & I \\
$^{195}$Hg$^{\rm g}$  &  49.0 m & \multicolumn{2}{d}{< 2.8} & \multicolumn{1}{c}{$\text{---}$}  & I & &
$^{151}$Tb        &  17.6 h &  3.55 &  0.20 &  3.55                                     & C \\
$^{191}$Au        &   3.2 h &  6.7  &  0.4  &  6.8                                      & I & &
$^{151}$Gd        &   124 d &  0.27 &  0.08 &  0.24                                     & I \\
$^{191}$Pt        &   2.9 d &  7.9  &  0.8  &  8.6                                      & I & &
$^{150}$Dy        &   7.2 m &  1.95 &  0.34 &  1.95                                     & C \\
$^{190}$Hg        &  20.0 m & \multicolumn{2}{d}{< 2.2} & \multicolumn{1}{c}{$\text{---}$}      & I & &
$^{150}$Tb$^{\rm g}$  &   3.5 h &  2.85 &  0.27 &  (2.9)                                    & C \\
$^{190}$Au        &  42.8 m &  5.0  &  0.4  &  5.0                                      & I & &
$^{149}$Gd        &   9.3 d &  3.12 &  0.20 &  3.13                                     & C \\
$^{190}$Ir$^{\rm m2}$ &   3.3 h &  2.04 &  0.15 & \multirow{2}[1]{5mm}[4mm]{ \[ \Bigr\} \; 4_{\cdot}0 \] } & I & &
$^{148}$Tb$^{\rm g}$  &   1.0 h &  1.13 &  0.16 &  (1.73)                                   & C \\
$^{190}$Ir        &  11.8 d &  1.95 &  0.10  &                                          & I & &
$^{147}$Gd        &  38.1 h &  2.9  &  0.2  &  2.9                                      & C \\
$^{189}$Pt        &  10.9 h & 12.5  &  1.1  & 12.2                                      & C & &
$^{147}$Eu        &  24.1 d &  0.44 &  0.20 &  0.51                                     & I \\
$^{189}$Ir        &  13.2 d &  5.1  &  0.8  &  4.8                                      & I & &
$^{146}$Gd        &  48.3 d &  2.24 &  0.13 &  2.23                                     & C \\
$^{188}$Pt        &  10.2 d & 13.2  &  0.4  & 10.6                                      & C & &
$^{146}$Eu        &   4.6 d &  0.73 &  0.08 &  0.71                                     & C \\
$^{188}$Ir        &  41.5 h &  5.6  &  0.4  &  5.8                                      & I & &
$^{145}$Eu        &   5.9 d &  2.15 &  0.34 &  2.19                                     & C \\
$^{187}$Pt        &   2.4 h &  9.8  &  1.5  & 10.2                                      & C & &
$^{143}$Pm        &   265 d &  2.0  &  0.4  &  1.9                                      & C \\
$^{187}$Ir        &  10.5 h &  7.3  &  1.0  &  6.8                                      & I & &
$^{139}$Ce        &   138 d &  1.2  &  0.1  &  1.2                                      & C \\
$^{186}$Pt        &   2.0 h & 10.4  &  1.8  &  8.8                                      & C & &
$^{135}$Ce        &  17.7 h &  0.81 &  0.21 &  0.78                                     & C \\
$^{186}$Ir$^{\rm m}$ &   2.0 h & \multicolumn{2}{d}{< 4.0}
                                            & \multirow{2}[1]{5mm}[4mm]{ \[ \Bigr\} \; 7_{\cdot}8 \] } & I & &
$^{132}$La        &   4.8 h &  0.32 &  0.08 &  0.34                                     & I \\
$^{186}$Ir$^{\rm g}$ &  16.6 h &  7.8  &  0.8  &                                           & I & &
$^{131}$Ba        &  11.5 d &  0.63 &  0.24 &  0.62                                     & C \\
$^{185}$Ir        &  14.4 h & 16.3  &  1.1  & 16.5                                      & C & &
$^{129}$Cs        &  32.1 h &  0.60 &  0.20 &  0.56                                     & I \\
$^{185}$Os        &  93.6 d &  5.2  &  0.8  &  4.1                                      & I & &
$^{127}$Xe        &  36.4 d &  0.51 &  0.05 &  0.52                                     & C \\
$^{184}$Pt        &  17.3 m &  8.2  &  0.8  &  9.0                                      & C & &
$^{124}$I         &   4.2 d &  0.16 &  0.07 &  0.15                                     & I \\
$^{184}$Ir        &   3.1 h &  8.3  &  0.5  &  9.6                                      & C & &
$^{121}$Te        &  16.8 d &  0.39 &  0.11 &  0.38                                     & C \\
$^{183}$Ir        &  58.0 m & 14.2  &  2.3  & 16.4                                      & C & &
$^{103}$Ru        &  39.3 d &  0.48 &  0.07 &  0.48                                     & C \\
$^{183}$Os$^{\rm m}$ &   9.9 h & 11.4  &  0.8  & \multirow{2}[1]{6mm}[4mm]{ \[ \Bigr\} \; 23_{\cdot}0 \] } & C & &
$^{96}$Tc         &   4.3 d &  0.23 &  0.05 &  0.23                                     & I \\
$^{183}$Os$^{\rm g}$ &  13.0 h &  9.7  &  0.6  &                                           & C & &
$^{95}$Nb         &  35.0 d &  0.65 &  0.07 &  0.65                                     & I \\
$^{183}$Re        &  70.0 d &  2.22 &  0.54 &  2.23                                     & I & &
$^{95}$Zr         &  64.0 d &  0.16 &  0.03 &  0.16                                     & C \\
$^{182}$Ir        &  15.0 m & 11.8  &  0.6  & \quad 14.8                                & C & &
$^{93}$Mo$^{\rm m}$   &   6.9 h &  0.23 &  0.04 & (0.23)                                    & I \\
$^{182}$Os        &  22.1 h &  9.8  &  1.0  &  9.9                                      & I & &
$^{89}$Zr         &   3.3 d &  0.58 &  0.05 &  0.59                                     & C \\
$^{181}$Re        &  19.9 h & 24.9  &  1.4  & 25.5                                      & C & &
$^{88}$Zr         &  83.4 d &  0.30 &  0.04 &  0.30                                     & C \\
$^{180}$Os        &  21.7 m & 19.9  &  0.8  & 19.9                                      & C & &
$^{88}$Y          &   107 d &  0.81 &  0.11 &  0.83                                     & I \\
$^{179}$Re        &  19.5 m & 20.8  &  2.0  & 20.3                                      & C & &
$^{87}$Y          &   3.3 d &  0.75 &  0.04 &  0.74                                     & C \\ 
$^{178}$Re        &  13.2 m & 15.5  &  2.4  & 15.7                                      & C & &
$^{86}$Y          &  14.7 h &  0.38 &  0.04 &  0.39                                     & C \\
$^{177}$W         &   2.3 h & 20.8  &  1.2  & 21.2                                      & C & &
$^{85}$Sr         &  64.8 d &  0.81 &  0.11 &  0.83                                     & C \\
$^{177}$Ta        &  56.6 h &  3.4  &  1.1  &  2.1                                      & I & &
$^{84}$Rb         &  32.8 d &  0.78 &  0.12 &  0.81                                     & I \\
$^{176}$Ta        &   8.1 h & 26.5  &  2.0  & 26.4                                      & C & &
$^{83}$Rb         &  86.2 d &  0.89 &  0.16 &  0.85                                     & C \\
$^{175}$Ta        &  10.5 h & 21.2  &  1.5  & 21.6                                      & C & &
$^{82}$Rb$^{\rm m}$   &   6.5 h &  0.40 &  0.11 & (0.43)                                    & I \\
$^{175}$Hf        &  70.0 d &  2.3  &  0.6  &  1.6                                      & I & &
$^{82}$Br         &  35.3 h &  0.31 &  0.08 &  0.30                                     & I \\
$^{173}$Ta        &   3.1 h & 16.1  &  2.7  & 15.5                                      & C & &
$^{75}$Se         &   112 d &  0.36 &  0.05 &  0.36                                     & C \\
$^{173}$Hf        &  23.6 h &  5.8  &  0.7  &  5.7                                      & I & &
$^{74}$As         &  17.8 d &  0.51 &  0.08 &  0.51                                     & I \\
$^{173}$Lu        &  1.37 y &  0.54 &  0.13 &  1.3                                      & I & &
$^{72}$As         &  26.0 h &  0.22 &  0.07 &  0.22                                     & I \\
$^{172}$Ta        &  36.8 m &  8.1  &  0.7  &  8.1                                      & C & &
$^{72}$Ga         &  14.1 h &  0.27 &  0.07 &  0.27                                     & I \\
$^{172}$Hf        &   1.9 y & 12.5  &  1.4  & 12.3                                      & I & &
$^{69}$Zn$^{\rm m}$   &  13.8 h &  0.34 &  0.04 & (0.32)                                    & I \\
$^{172}$Lu        &   6.7 d &  1.34 &  0.20 &  1.36                                     & I & &
$^{59}$Fe         &  44.5 d &  0.36 &  0.07 &  0.38                                     & C \\
$^{171}$Lu        &  8.24 d & 20.8  &  1.0  & 20.9                                      & C & &
$^{48}$V          &  16.0 d &  0.19 &  0.03 &  0.19                                     & C \\
$^{170}$Hf        &  16.0 h & 16.1  &  1.4  & 16.1                                      & C & &
$^{46}$Sc         &  83.8 d &  0.24 &  0.07 &  0.24                                     & I \\
$^{170}$Lu        &   2.0 d &  3.8  &  0.7  &  3.9                                      & I & &
$^{41}$Ar         &   1.8 h &  0.22 &  0.05 &  0.22                                     & C \\
$^{169}$Lu        &  34.1 h & 16.0  &  0.6  & 16.0                                      & C & &
$^{24}$Na         &  15.0 h &  0.27 &  0.07 &  0.27                                     & C \\
$^{169}$Yb        &  32.0 d &  2.15 &  0.67 &  2.18                                     & I & &
$^{22}$Na         &   2.6 y & \multicolumn{2}{d}{< 2.7} & 0.31                        & C \\
$^{166}$Yb        &  56.7 h & 15.5  &  0.8  & 15.5                                      & C & &
$^{7}$Be          &  53.3 d & \multicolumn{2}{d}{< 3.5} & \multicolumn{1}{c}{$\text{---}$}                  & C \\
\colrule
\end{longtable*}
\endgroup

\section{DATA ANALYSIS}
\label{sec:analysis}

The method based on off-line $\gamma$-ray spectroscopy has some limitations. The most important is the 
necessity to use some phenomenological model to reconstruct the yields of unobservable products 
\cite{Rudstam66,Silberberg73,Moser89,Summerer90,Egidy91}. In similar experiments with other projectiles
the data are sparsely spread over the $N$-$Z$ plane. In this case less detailed models may be applied 
and the results obtained are only a first order approximation of the true yield distribution $Y(A,Z)$, 
even when cumulative cross sections are involved in the fitting procedure. Moreover, the precision of 
tabulated absolute $\gamma$-ray intensities sometimes leaves much to be desired and, finally, the 
decomposition of spectra with a few hundreds of lines, as they are measured for heavier targets, becomes 
a challenge for the persistence of the evaluator.

Targets irradiated with antiproton beams, much less intense than the proton beams, have a rather low 
activity level. In particular, yields obtained for the fission fragments were close to our detection limit. 
Keeping in mind one of our goals, the estimation of the probability for antiproton induced fission, 
the data analysis needed special care and some feedback. A primary set of yield results obtained from 
the spectrum analysis served as an input for the model distribution fitting at its early stage, when 
the best approach was searched for. This relates to the choice of the final formula as well as to the 
division of the data to subsets assuring the lowest total $\chi^{2}$. Afterwards, a modeling procedure 
was applied to check, confirm or eliminate some doubtful experimental yields. For some mass regions it 
appeared necessary to apply an additional or separate evaluation, and we describe it at the end of this section. 

\subsection{Fitting procedure}
\label{sec:procedure}

The formula, used to describe the yield distribution was rather complex in order to be as universal as possible 
and to test various models. This complexity mainly arose from the aim of taking into account cumulative yields 
and from introducing the ($N,Z$) evenness corrections. The general formula was factorized into two components, 
mass and charge distributions, $\rm Y_{A}$ and $\rm Y_{ZP}$, respectively,
\begin{equation}
Y(A,Z) = Y_{A} \times Y_{ZP}.
\label{general}
\end{equation}

The distribution over the mass (the main distribution ridge) was modeled with the exponential of a fourth-order 
polynomial with parameters $a_{1}-a_{5}$,
\begin{equation}
Y_{A} = e^{(a_{1}+a_{2}A+a_{3}A^{2}+a_{4}A^{3}+a_{5}A^{4})}.
\label{mass}
\end{equation}
This was useful for testing the fit in broader mass regions, where the ridge shape may change more
rapidly.

The form of the second factor in Eq.\ (\ref{general}), $Y_{ZP}$, (the charge distribution) 
was multiplied by the odd-even corrections $F_{P}$. When needed, an additional component,
containing the sum of yields of the decay predecessors of given ($A,Z$) isotope, was added here
\begin{equation}
Y_{ZP}=\sum_{k=0}^{5} F_{P}(Z+ck,N-ck) \times e^{-(Z+ck-Z_{p})^{w}/2\sigma^{2}},
\label{charge}
\end{equation}
where the term with $k = 0$ corresponds to the independent $Y_{ZP}(A,Z)$ yield and the terms with $k = 1\div 5$ 
stand for the precursors contributions. The upper limit of the sum over $k$ was set to 5, because the
charge distribution for given $A$ is rather narrow and neglecting $k > 5$ did not change
the sum by more than 1\%.

The most probable charge path $Z_{p}$ and the charge distribution width $\sigma$ were expressed as
a third-order polynomials of $A$, with parameters $a_{6}-a_{9}$ and $a_{10}-a_{13}$,
respectively,
\begin{equation}
Z_{p} = a_{6}+a_{7}A+a_{8}A^{2}+a_{9}A^{3},
\end{equation}
\begin{equation}
\sigma = a_{10}+a_{11}A+a_{12}A^{2}+a_{13}A^{3}.
\end{equation}

The value of the factor $c$ in the sum of the yield cumulation for a given $A$ depended on the side
of the stability valley on which the given isotope lies,
\begin{equation}
c =
\begin{cases}
1,& \text{EC, } \beta^{+} \text{decay (neutron-deficient nuclei)};\\
-1,&  \beta^{-} \text{decay (neutron-rich nuclei).}
\end{cases}
\end{equation}

The power index $w$ in the exponent argument in Eq.\ (\ref{charge}) was allowed to be different for 
$Z > Z_{p}$ and $Z \le Z_{p}$,
\begin{equation}
w =
\begin{cases}
a_{16},& Z > Z_{p}\\
a_{17},& Z \le Z_{p},
\end{cases}
\end{equation}
where $a_{16}$ was always set to 2 and $a_{17}$ = 2 or $a_{17}$ = 1.5 was used in order to test the asymmetric 
charge distribution in the latter case.

Finally, the odd-even correction was assumed to be a simple factor depending on the $Z$ and $N$ evenness
combination
\begin{equation}
F_{P} =
\begin{cases}
1,& \text{Z - even, N - even};\\
a_{14},& \text{Z - odd, N - even};\\
a_{15},& \text{Z - even, N - odd};\\
a_{14}a_{15},&  \text{Z - odd, N - odd.}
\end{cases}
\label{pairing}
\end{equation}

\begin{figure}
\includegraphics[width=\columnwidth]{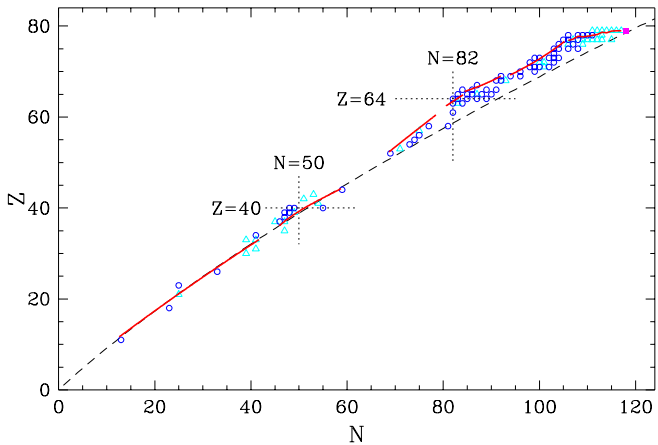}
\caption{\label{fig:aust_map}The path of the most probable atomic number $Z_{p}$ corresponding to the best 
fit values for the parameters $a_{6}-a_{8}$
in seven mass regions (solid lines). Positions of nuclei, for which a cumulative or an
independent yield was determined, are denoted by circles and triangles, respectively. Dashed line
- valley of stability approximated by the relation $Z_{\beta } = A/(1.98+0.0155A^{2/3})$
\protect\cite{Marmier71}. Open square indicates the target nucleus, vertical and horizontal dotted
crosses are plotted for the magic $N = 82$ and the closed shell $Z = 64$ and for the magic $N = 50$ 
and the closed shell $Z = 40$.}
\label{aust_map}
\end{figure}

A division of the whole $A$-$Z$ plane into subregions may be treated as yet another model parameter. 
To have control over it we have plotted positions of all detected nuclei in the $N$-$Z$ plane; this 
appeared to be very helpful for a preliminary determination of the $Z_{p}$ path, especially  for regions 
with many data. The final mass region division is illustrated in Fig.~\ref{aust_map} by the solid lines
showing $Z_{p}$ fitted for seven data regions. Mass range limits were fixed to get the smallest total 
$\chi ^{2}$ for the whole data range and to possibly simplify the model for the course of the $Z_{p}$ path. 
We have tested many alternative divisions, especially for the region of the heavy evaporation residues 
($143 \leq A \leq 183$). The fitting applied to broader $A$ ranges than those listed in Table \ref{parameters} 
resulted 
in at least one order of magnitude larger $\chi ^{2}$ values, mainly due to rapid changes in the $Z_{p}$ 
course at $A$ = 162 and $A$ = 150. For three separate regions of the lighter products with $A$ = 121-139, 
82-103 and 24-75 the limits were defined by the grouping of the experimental data. 

As may be seen from Table \ref{parameters}, presenting the final parameters, sometimes the best fit is 
obtained when the number of parameters exceeds the number of data. This was done by fixing some parameters when 
the others were fitted, and vice versa. Various combinations and order of fixing/releasing of parameters 
as well as their total number were tested. The shape of the mass ridge could be parametrized with a maximum 
of four parameters, the most probable $Z$ path was approximated via a parabola except for two cases and the
charge distribution width was constant or changed linearly with mass. Odd-even corrections were applied only 
for three heaviest mass regions, where a larger number of points and smaller relative errors of the
experimental data allowed to get a reliable fit. The shape of the charge distributions was modeled better 
with the use of an asymmetric form for the evaporation residues, lying further from the stability valley 
(see Fig.~\ref{aust_map}). For lighter, fission products the $Z_{p}$ path goes closely along the valley of
stability and here a symmetric Gaussian shape was more adequate.

\begingroup
\begin{table*}
\caption{Best fit parameters obtained for the yield distribution model applied to six mass regions.}
\label{parameters}
\begin{ruledtabular}
\begin{tabular}{rrrrrrr}
Paramater & \multicolumn{6}{c}{Mass range} \\
& \multicolumn{1}{r}{163-182} & \multicolumn{1}{r}{150-161} & \multicolumn{1}{r}{143-149} 
& \multicolumn{1}{r}{121-139} & \multicolumn{1}{r}{82-103} & \multicolumn{1}{r}{24-75} \\
\colrule
$Y_{A} \quad a_{1}$  & -93.98(2)   & -90.63(3)   & -479.2(1)    & 18.13(9)     & -30.97(5)   & -2.07(11)   \\
$a_{2}$ & 1.0951(1)   &  1.060 (2)  & 6.486(1)     & -0.207(1)    & 0.690(1)    & -0.0059(18) \\
$a_{3}$ & -0.00310(1) & -0.00302(2) & -0.02192(3)  & -0.000706(5) & -0.00390(6) &  0.00029(3) \\
$a_{4}$ &             &             & -0.3(2)E-7   &  0.884(4)E-5 &             &             \\
$Z_{p}$ \quad $a_{6}$ & 96.26(4)    & 136.6(1)    & -3.06(9)     & -7.2(3)      & -142.1(1)   &  0.77(10)   \\
$a_{7}$ & -0.6681(2)  & -1.201(1)   & 0.5051(6)    &  0.525(3)    &  4.886(1)   &  0.456(2)   \\
$a_{8}$ & 0.003081(1) & 0.00485(5)  & -0.000329(4) & -0.00029(2)  & -0.0456(1)  & -0.000329(2)   \\
$a_{9}$ &              &             &              &  0.8(14)E-7  & 0.000153(1) &             \\
$\sigma$ \quad $a_{10}$ & -0.82(2)    & 1.17(4)     & 1.12(8)      & 0.91(9)      & 1.15(4)     &  2.12(11)   \\
$a_{11}$ & 0.0105(2)   &             &              &              &             & -0.013(2)   \\
$F_{P}$ \quad $a_{14}$ & 0.67(3)     & 0.82(5)     & 0.58(7)      & 1.           & 1.          & 1.          \\
$a_{15}$ & 0.74(3)     & 0.91(6)     & 0.90(8)      & 1.           & 1.          & 1.          \\
$w$ \quad $a_{16}$ & 2.          & 2.          & 2.           & 2.           & 2.          & 2.          \\
$a_{17}$ & 1.5         & 1.5         & 1.5          & 1.5          & 2.          & 2.          \\
$\chi^{2}/NDF$ & 0.045 & 0.006       & 0.164        & 0.131        & 0.082       & 0.020       \\
\end{tabular}
\end{ruledtabular}
\end{table*}
\endgroup

\begin{figure}[b]
\begin{center}
\includegraphics[width=7cm]{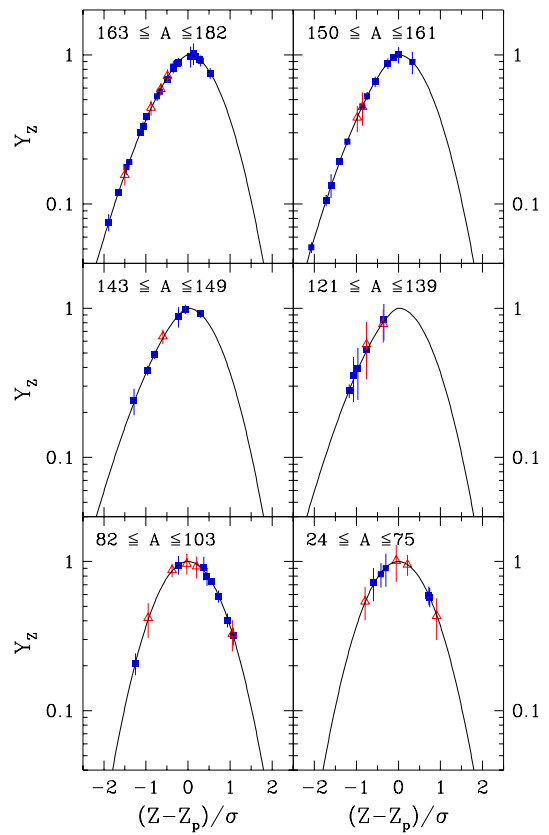}
\end{center}
\caption{Normalized Z distributions for six mass ranges. The fitted function (line)
is the exponent $\exp(-|(Z-Z_{p})/\sigma |^{w})$, where $w$ is the charge distribution power
index, $Z_{p}$ the most probable $Z$ and $\sigma $ the Z distribution width. Experimental
values ($\triangle$ - independent, $\blacksquare$ - cumulative) are normalized as described
in the text. All distributions have the same normalized width equal to unity (please note the asymmetric 
shapes for four distributions of heavier isotopes, with different slope for both sides).}
\label{aust_cha}
\end{figure}

The normalized charge distributions for six mass intervals are plotted in Fig.~\ref{aust_cha} against 
the normalized charge difference $(Z-Z_{p})/\sigma$. This reduces the distributions to the same width in the 
case where $\sigma$ is not constant in the given region. The fitted function is the simple exponent 
$\exp(-|(Z-Z_{p})/\sigma|^{w})$, then, for comparison, experimental data $Y(A,Z)_{E}$ are normalized 
with three factors coming from the fit
\begin{equation}
Y_{Z}=\frac{Y(A,Z)_{E}}{Y(A)\times F_{P}(N,Z) \times f_{I}}.
\end{equation}
Here, the mass distribution $Y(A)$ is as in Eq.\ (\ref{general}), the odd-even correction $F_{P}(N,Z)$
follows Eq.\ (\ref{pairing}) and the factor $f_{I}$ corresponds to the independent yield fraction in
the case of the cumulative yield
\begin{widetext}
\begin{equation}
f_{I}=\frac{\D F_{P}(Z,N)\int_{Z-0.5}^{Z+0.5} e^{-(z-Z_{p})^{w}/2\sigma^{2}}dz}
{\D \sum_{k=0}^{5}\big[F_{P}(Z+ck,N-ck)\times \int_{Z+ck-0.5}^{Z+ck+0.5} e^{-(z+ck-Z_{p})^{w}/2\sigma^{2}}dz\big]}.
\end{equation}
\end{widetext}
Sometimes charge distributions are normalized to unity integral over Z to get the total yield for a given A
equal directly to $Y_{A}$ \cite{Summerer90,Sihver92}. However, when odd-even corrections are used the
distribution of the total yield cannot be described by a simple continuous function as in
Eq.\ (\ref{mass}). Also the generalization of the charge distribution shape with the two-valued (or released)
index $w$ leads to problems in obtaining an analytical form of the normalization factor for this
function. Hence our distributions for six mass regions are normalized only to the same width $\sigma$ = 1, 
not to the same integral.

\subsection{Treatment of the heaviest residues}
\label{sec:heaviest}

\begin{figure}[h]
\begin{center}
\includegraphics[width=\columnwidth]{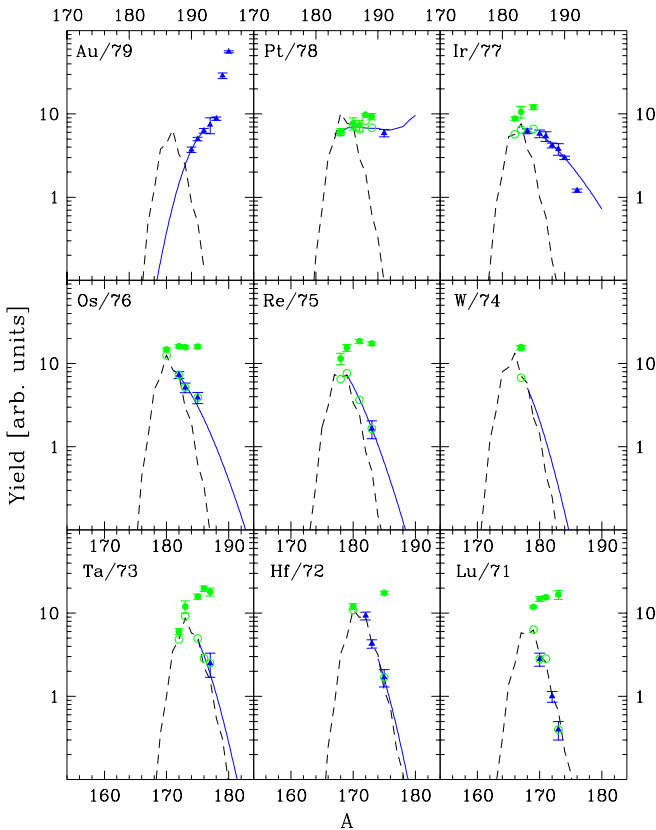}
\end{center}
\caption{Isotopic yield distributions for the nine heaviest residual elements produced in the fragmentation
of gold with stopped antiprotons. Dashed lines show yields obtained from the fit applied to the region
$163 \le A \le 182$;
triangles, full and open circles represent the experimental yields: independent, cumulative and cumulative 
transformed to independent, respectively. Solid lines illustrate the corrections of the isotopic yield 
distribution for the heaviest isotopes of the elements from Pt to Hf, and for the lightest Au isotopes 
(see Sec.\ \ref{sec:analysis} for discussion).}
\label{aust_iso}
\end{figure}

It is a well known fact that none simple phenomenological model can properly describe the charge 
distribution of yields for isotopes lying near the target nuclide \cite{Silberberg73,Summerer90}. 
The main reason for this is the asymmetric, non-Gaussian shape of the charge distribution, with the most 
probable $Z$ and width rapidly changing with mass. Such a phenomenon is clearly seen when one uses
a longer section of the $Y(A,Z)$ yield, along constant $Z$ value instead of constant $A$. Fig.~\ref{aust_iso}
presents isotopic $Y_{A}(A)$ distributions obtained in this experiment for elements ranging from Au to Lu. 
Also isotopic distributions obtained for the heaviest element after stopped antiproton absorption on $^{176}$Yb,
$^{148}$Nd and $^{130}$Te targets \cite{Lubinski97} exhibit such a behavior: a steep and narrow
distribution for the target element $Z_{T}$, a flat and broad distribution for the $Z_{T}$-1 element and deformed
quasi-Gaussian shapes for some smaller $Z$, with the deformation on the heavy mass side decreasing
with increasing distance from $Z_{T}$. Even though the low mass side for all elements but
$Z_{T}$ may be described with the same slope, the slope at the higher mass side changes rapidly and cannot
be fitted well with a fixed isotopic distribution asymmetry, i.e. with the unique, constant $a_{17}$ parameter. 
As a consequence, the heaviest elements should be excluded from the global fit and their $Y_{A}(A)$ yields 
have to be fitted separately for a given $Z$.

\begin{figure}[t]
\begin{center}
\includegraphics[width=\columnwidth]{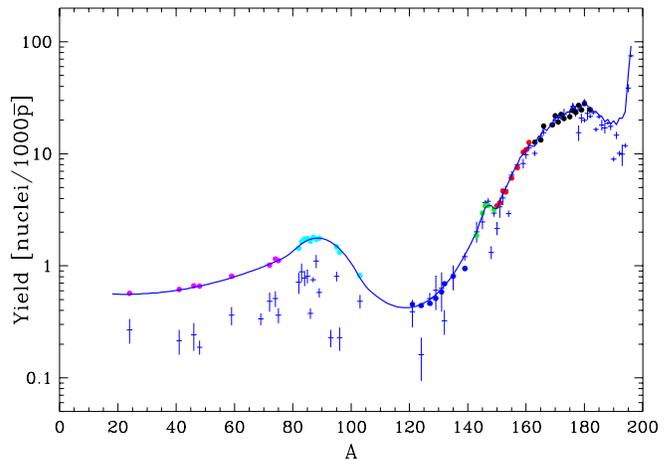}
\end{center}
\caption{Mass distribution of residual nuclei produced by the reaction of stopped $\rm \overline p$ with Au.
The cumulative sum of experimentally observed yields for a given mass is shown with crosses,
the corresponding fitted $Y(A)$ yield for all input data (averaged over $Z$ for given $A$) is represented
by full circles. The line shows the sum of fitted yields completed with interpolated yields for
mass intervals not fitted. For the heaviest isotopes ($A \ge 175$) an additional correction of the mass
yield data was performed, as described in Sec.\ \ref{sec:analysis}.}
\label{aust_sum}
\end{figure}

The method of the yield completion for the heaviest elements is recursive: at the beginning we estimate the
lacking yields of the lighter Au isotopes. With the use of these results cumulative, experimental yields 
for Pt are converted to independent ones and the isotopic distribution for this element is evaluated.
Then, a similar procedure is applied for Ir, Os and so on. The method was applied down to Ta and Hf elements, 
where the yields for $A \ge 175$ were corrected. Finally, the summed $Y(A)$ yields for $A \ge 175$ presented 
with a line in Fig. \ref{aust_sum} are the combination of results of both evaluations: the $163 \le A \le 182$ 
region global fit and the procedure described above.

The platinum distribution was the most laborious case, due to lack of radioactive isotopes above mass 191 
and owing to the strong odd-even $N$ effect (up to 30\%), observed for this even-$Z$ element produced after 
low-energy absorption of antiprotons. The overall shape of the isotopic distribution was assumed 
to be similar to that observed for Tm residues after Yb fragmentation with antiprotons \cite{Lubinski97}, 
with an increasing enhancement of yields for three heaviest nuclei, which for Pt are these with the 
mass numbers 194, 195 and 196. 
The number for $^{196}$Pt obtained in this way ($16/1000\rm\overline p$) was confronted with the result 
of another estimation, based on the so-called halo factor dependence on separation energy $S_{n}$ of the
neutron from target nucleus \cite{Lubinski98}. For heavy nuclei with $S_{n}$ close to 8 MeV the halo factor 
is of the
order of $4 \div 5$, hence, using the $^{196}$Au yield the $Y(^{196}$Pt) should be between 14 and 23 
per 1000$\rm \overline p$. These two estimations are consistent. For the lightest Pt isotopes we assumed 
that the steep slope coming from the $163 \le A \le 182$ region fit is a good approximation, this assumption was 
used also for consecutive, lower $Z$ elements (we have checked the justification of this approach using the 
lightest mass data points of these isotopes in an additional test fit). At last, on the basis of observed changes
of Pt yield for odd and even $N$ isotopes an appropriate correction was applied.

\subsection{Region of the $\alpha$-decay}
\label{sec:alpha}

There is a narrow bump at $A$ = 147 in the mass yield distribution, a feature observed for Au 
\cite{Kaufman80,Michel97} and for Ta \cite{Kozma91,Gulda93} target fragmentation after reactions with 
protons, heavy ions and stopped antiprotons. The enhancement of the cross section in this region was suggested 
to be the result of $\alpha$-decay of nuclei above the $N$ = 82 shell \cite{Kaufman80}, but the authors abstained 
to estimate quantitatively this effect due to lack of charge dispersion curves which could not be fitted for
limited experimental data. We have done such an estimation for our data performing a preliminary fit for 
isotopes 
not affected by $\alpha$-decay, i.e. for $154 \le A \le 161$, adding $^{150}$Dy with an independent yield. 
Taking into account a charge dispersion yield $Y(A,Z)$ obtained in this way we have calculated 
appropriate decay corrections for experimentally measured yields of $^{153}$Tb (+2.9\% correction), 
$^{153}$Gd (+2.8\%), $^{152}$Dy (+7.2\%), $^{152}$Tb (+6.0\%), $^{151}$Tb (+14.8\%), $^{151}$Gd (+13.6\%), 
$^{150}$Tb (+32.5\%), $^{149}$Gd (+5.5\%), $^{147}$Gd (-13.7\%), $^{147}$Eu (-12.1\%) and $^{146}$Gd (-23.7\%).
The experimental results listed in Table \ref{yields} are the corrected ones and were used for fitting in two
mass ranges affected by this effect. As can be seen from Fig. \ref{aust_sum} the corrections obtained are too 
small to remove the local yield maximum at $A$ = 147 (crosses show the yield before correction, circles after 
that). Therefore, $\alpha$-decay alone cannot explain fully such a feature and the observed yield enhancement
in this region should be partially ascribed to the closed $N$ = 82 shell influence.

\section{RESULTS}
\label{sec:results}

The experimental data are presented in Table \ref{yields}, together with the fit results for isotopes
representing full yield for a given pair of $A$ and $Z$. Results are normalized to yield per 1000 
$\rm \overline{p}$ with the total number of antiprotons stopped in the target ($9.25\times 10^{8})$.
The final mass yield distribution is presented in Fig. \ref{aust_sum}. The cumulative sum of all yields
observed for a given mass number $A$ is here compared with the total yield $Y(A)$ obtained from the fit via 
summation of all fitted $Y(A,Z)$ values over $Z$, or from the interpolation between fitted mass regions. 
The global curve of the fitted $Y(A)$ yield, when compared with the summed
experimental yields, forms its exact skyline in almost the whole region of the evaporation residues. A
deviation from this rule is observed for three mass ranges: the heaviest, with $A > 176$, a few mass 
numbers around $A$ = 147 and all fission fragments ($A < 120$). Except for the second region (affected by 
the $\alpha$-decay), this is the result of prevailing accumulation of the isobaric yield by non-detectable 
isotopes. The depression of observed yield of the heaviest evaporation residues is 
narrow but deep, with a maximum decrease to about 40\% of the fitted $Y(A)$ for $A \approx 192$, and comes
mainly from the stable Pt isotopes produced. For fission products, where the $Z_{p}$ path 
goes over the stability valley, the observed yield is strongly suppressed and its outline reaches only 
about 20-50\% of the fitted yield.  

Leaving out two heaviest masses, the maximal $Y(A)$ yield is reached at mass 180 but the
largest individual $Y(A,Z)$ production is fitted for $^{176}$W. The small yield peaks observed
for some even masses ($A$ = 180, 176, 170, ...) are due to the strongest odd-even effect for
some even $Z$ lying almost on the $Z_{p}$ path. On the other hand, the global
mass yield minimum appears between $A = 105$ and $A = 120$. As numerous $\gamma$-lines of strongly 
populated heavier nuclei covered this region, no valuable production limits can be given for
this region and we have to stick to the interpolated curve.

After evaluation of the mass yield curve it is possible to estimate the relative yields for different reaction 
channels. The fission fragments mass range should be treated with some care as their multiplicity 
is equal to 2 or greater when one takes into account any multifragmentation process. Assuming that 
all residues with $40 \leq A \leq 120$ are binary fission products (i.e. two heavy residues 
per antiproton) and neglecting the lightest $A < 40$ masses, we have obtained the summed fission 
yield. Comparing this number with the total $Y(A)$ integral in the mass limits from 40 to 196 we have
extracted the probability of gold fission induced with stopped antiprotons to be ($3.8\pm 0.5$)\%. 
The lighter mass region ($A = 10-40$), not taken into account in fission due to possible multiplicity $> 2$  
and/or the not fully negligible chance to have a fission partner in the $A > 120$ region, constitutes 
additionally less than 0.9\% of the total yield (the error quoted above takes this into account). Our result 
compares well with the fission probability of 3.1(3)\% obtained in an experiment where fission fragments 
yields were measured with PIN diodes \cite{Schmid94} and is substantially larger then the 
value of 1.5\% derived from another experiment using also on-line technique \cite{Bocquet92}.

\section{DISCUSSION}
\label{sec:discussion}

\subsection{Antiprotons versus other projectiles}
\label{sec:projectiles}

The properties of reactions induced by stopped antiproton absorption can be investigated
by comparison with yield distributions obtained for other, more "classical" projectiles.
We have confined this comparison to the gold target as the literature is quite rich here 
\cite{Kaufman80,Michel97,Summerer90,Sihver92,Kaufman76,Kaufman79,Mustapha99}. The two other popular 
neighbor-mass targets, Pb and Ta, represent rather different decay scenarios, with respectively more 
and less pronounced fission channel.

\subsubsection{Mass yield curve}

\begin{figure}[t]
\begin{center}
\includegraphics[width=\columnwidth]{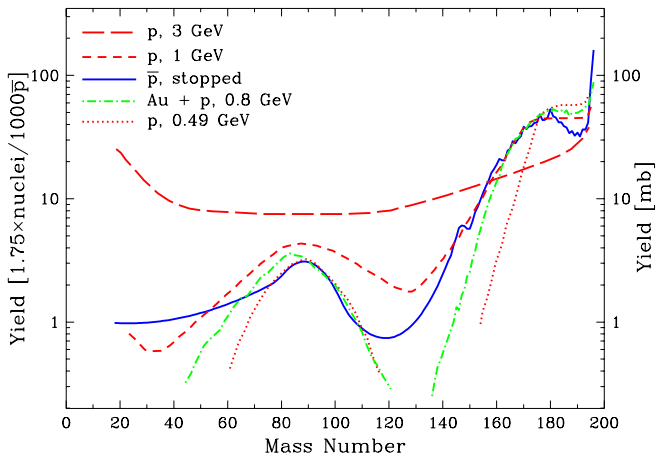}
\end{center}
\caption{Comparison of the mass yield distribution obtained for protons and stopped antiprotons.
Curves for protons at 0.49 GeV, 1.0 GeV and 3.0 GeV adopted from \protect\cite{Kaufman80}, for the inverse 
kinematic reaction of Au on H at 800 MeV from \protect\cite{Rejmund01,Benlliure01}. The yield axis for 
antiprotons was normalized with a factor of 1.75.}
\label{aust_com}
\end{figure}

First, we present a rather qualitative comparison with the yield curve shapes extracted for protons.
Figure \ref{aust_com} shows the summed isobaric $Y(A)$ yield obtained for stopped antiprotons
plotted together with yield distributions resulting from Au fragmentation by 0.49, 0.8, 1.0 
and 3.0 GeV protons \cite{Kaufman80,Rejmund01,Benlliure01}. Since the yields for stopped antiprotons 
and protons are measured in different units, we have normalized our yield axis with an arbitrary 
factor equal to 0.57, providing the concordance between $\overline{p}$'s and 1 GeV protons results 
in the 150-170 mass range. It should be stated here that the yield curve presented for fission residues
in the case of 1 GeV protons was fitted with only 5 mass points \cite{Kaufman80}. 

The most striking differences between the curves shown in Fig.~\ref{aust_com} are seen for the 
fission region. The fission probability for gold excited by protons, estimated as in 
Sec.\ \ref{sec:results}, is equal to $\approx$ 6.5\%, $\approx$ 3.7\% and $\approx$ 3.3\%
for 1 GeV, 0.8 GeV and 0.49 GeV protons, respectively. Then, 800 MeV protons seem to correspond 
to stopped antiprotons, but fission takes only a small part of the total yield and the comparison 
of distribution shapes in the evaporation region is much more adquate. Such inspection leads to the 
conclusion that stopped antiprotons match with protons at 1 GeV. 

Besides the level and width of the fission hump, a second major feature distinguishing the mass yield 
shapes observed for stopped antiprotons and protons is the distribution for the heaviest masses close 
to the target. Here the experimental situation is much better than in the fission case: more reliable 
nuclear spectroscopy 
data is additionally confirmed by the results of inverse kinematic measurements. The yield distribution 
for protons is rather unchanged in energy range from 0.5 to 1 GeV and forms a plateau between 
$A \approx 175$ and $A \approx 194$. On the contrary, for antiprotons in this mass range, not only 
for the gold target \cite{Szmalc92,Gulda93,Lubinski97}, the yield slowly decreases 
from $A \approx A_{T}-20$ to $A \approx (A_{T}-3)$ and then strongly rises, reaching 
an absolute maximum at $A = (A_{T}-1)$. The enhancement of yield for few masses closest to the 
target may be explained by two mechanisms. The first one is the soft antiproton absorption, 
where almost all annihilation pions miss the rest of the target nucleus. Then the $(A_{T}-1)$ nuclei are 
left with very low excitation energy. Only after antiproton absorption on nucleons occupying 
a deeper states \cite{Wycech96} rearrangement of nucleon configurations results in a mass loss of one or 
two additional units. The probability of the production of ($A_{T}-1$) nuclei is quite large, about 10\% for targets 
used in our nuclear periphery studies \cite{Lubinski98,Schmidt99} and the results obtained for gold are 
also of this order of magnitude (cf. Table \ref{yields}). The second mechanism leading to low excitation 
energies
is the class of all processes where the annihilation pions escape unabsorbed by the target nucleus
but due to the sizeable total $\pi$-nucleus cross section excite this nucleus enough to emit few
nucleons. The probability of such kind of quasi-elastic meson escapes may be quite large when the nuclear
diffuseness and partial opacity are considered \cite{Cugnon01}.

\begin{figure}[t]
\begin{center}
\includegraphics[width=\columnwidth]{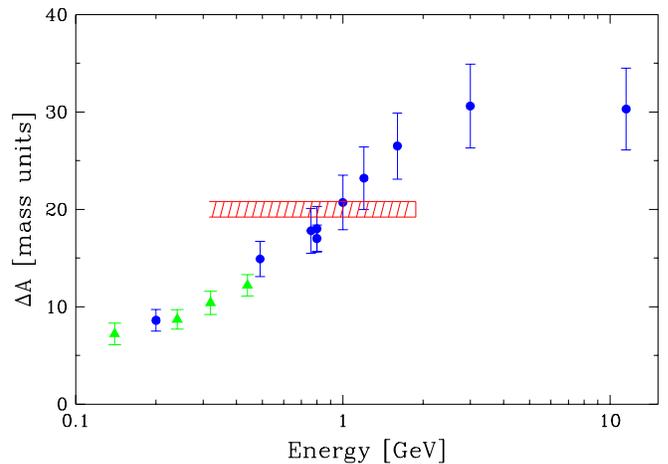}
\end{center}
\caption{Average mass removal from the Au targets irradiated with: $\blacktriangle$, pions; $\bullet$, protons 
(Refs. cited in Table \ref{projectiles}) and stopped antiprotons (pion data are shifted right by the $\pi$ rest 
mass). The hatched band shown for antiproton data reflects the error in $\Delta A$ and a possible range of 
energy deposition for stopped $\rm \overline{p}$ annihilation with one of the target nucleons.}
\label{aust_dep}
\end{figure}

A quantitative comparison between antiprotons and other projectiles is presented in Table 
\ref{projectiles} and Fig. \ref{aust_dep}. For this purpose we have calculated the average 
mass removed from the target, $\Delta A$, defined as: 
\begin{equation}
\Delta A=A_{T}^{\star}-\frac{\D \int_{A_{min}}^{A_{max}} Y(A)AdA}
{\D \int_{A_{min}}^{A_{max}} Y(A)dA},
\end{equation}
where $A_{T}^{\star}$ equals $A_{T}$ for protons and pions or $A_{T}-1$ for antiprotons.
$A_{\rm min}$ is the lower integration limit adjusted to get all single heavy residues
and $A_{\rm max}$ is equal to $A_{T}$ for protons and pions or to $A_{T}-2$ for antiprotons.
The residual mass $A_{T}-1$ is ignored in the integration of the reaction yields for
antiprotons since it attests no reaction (soft $\overline{p}$ absorption).

\begin{table}[b]
\caption{Characteristics of the mass yield distribution after the reaction of different projectiles 
with a gold nuclei.}
\label{projectiles}
\tabcolsep1pt
\begin{ruledtabular}
\begin{tabular}{cdcc}
Projectile     & \rm Energy               & $\Delta A$    & Ref.               \\
               & \rm [GeV]                &               &            \\
\colrule
$\pi ^{+,-}$   & 0.0                    & $\phantom{1}7.2\pm 1.1$  & \phantom{}\cite{Pruys79}  \\
$\pi ^{+,-}$   & 0.1                  & $\phantom{1}8.7\pm 1.0$  & \phantom{}\cite{Kaufman79}  \\
$\pi ^{+,-}$   & 0.18                 & $10.4\pm 1.2$ & \phantom{}\cite{Kaufman79}  \\
$\pi ^{+,-}$   & 0.3                  & $12.2\pm 1.3$ & \phantom{}\cite{Kaufman79}  \\
p              & 0.2                  & $\phantom{1}8.6\pm 1.1 $ & \phantom{}\cite{Kaufman80}  \\
p              & 0.49                 & $14.9\pm 1.8$ & \phantom{}\cite{Kaufman80}  \\
p              & 0.8\footnotemark[1] & $17.0\pm 1.4$ & \phantom{}\cite{Mustapha99} \\
p              & 0.76                 & $17.8\pm 2.3$ & \phantom{}\cite{Michel97}  \\
p              & 0.8                  & $18.0\pm 2.3$ & \phantom{}\cite{Michel97}  \\
p              & 1.0                  & $20.7\pm 2.8$ & \phantom{}\cite{Kaufman80} \\
p              & 1.2                  & $23.2\pm 3.2$ & \phantom{}\cite{Michel97}  \\
p              & 1.6                  & $26.5\pm 3.4$ & \phantom{}\cite{Michel97}  \\
p              & 2.6\footnotemark[2] & $30.1\pm 1.9$ & \phantom{}\cite{Summerer90} \\
p              & 3.0                  & $30.6\pm 4.3$ & \phantom{}\cite{Kaufman80} \\
p              & 11.5                 & $30.3\pm 4.2$ & \phantom{}\cite{Kaufman76} \\
p              & 800.0                & $26.1\pm 3.7$ & \phantom{}\cite{Sihver92}  \\
$\rm \overline p$  & 0.0\footnotemark[2] & $17.3\pm 1.3$ & \phantom{}\cite{Egidy91} \\
$\rm \overline p$  & 0.0              & $20.0\pm 0.8$ & this work \\
\end{tabular}
\end{ruledtabular}
\footnotetext[1]{Inverse kinematic reaction.}
\footnotetext[2]{Fit for generalized formula.}
\end{table}

To get consistent and comparable results, for each 
data set taken from the literature we have applied a uniform method to determine $\Delta A$. 
Only experimental data representing the highest (approximately the whole) cumulative yield for 
a given $A$ were used to construct the mass yield distribution curve. The absolute errors of the
quantities presented in Table \ref{projectiles} were estimated to be of the order of 10-20\%, but the 
relative errors should be smaller. In addition to proton data, results for pions absorbed by the gold 
target are presented, their energy range is limited as compared to the rest of data but coincides 
with the kinetic energy of pions emitted in antiproton annihilation. The average mass removal from 
the target 
nucleus smoothly correlates with the projectile energy. As can be seen from Table \ref{projectiles} 
and Fig. \ref{aust_dep}, the average mass removed from the gold target by stopped antiprotons is only 
slightly lower than $\Delta A$ obtained for 1 GeV protons. 

\subsubsection{Charge distribution}

\begin{figure}[t]
\begin{center}
\includegraphics[width=\columnwidth]{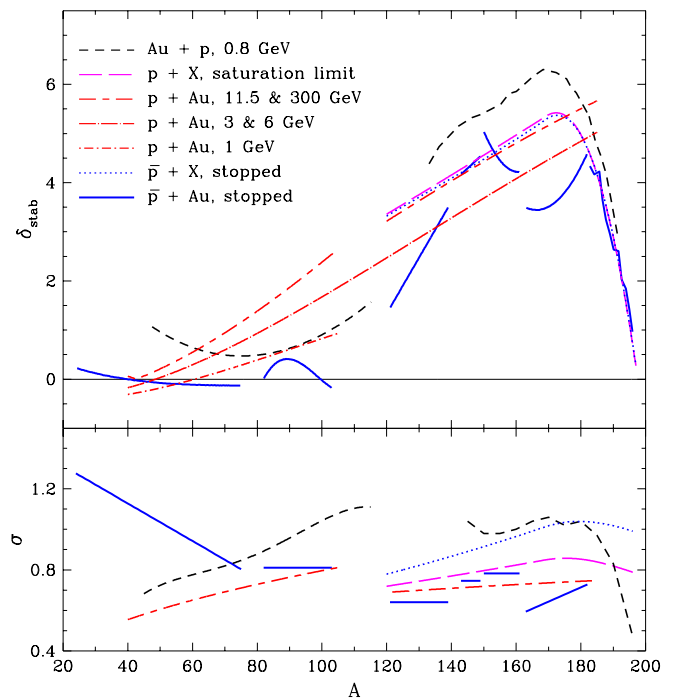}
\end{center}
\caption{Upper part: Distance of the fitted most probable Z from the beta stability line: 
protons/heavy ions compilation at saturation limit \protect\cite{Summerer90}, antiprotons at rest energy 
compilation \protect\cite{Egidy91}, Au at 800 MeV on H \protect\cite{Rejmund01,Benlliure01}, protons
at 11.5/300 GeV \protect\cite{Kaufman76}, protons at 3/6 GeV \protect\cite{Kaufman80} and fission for 
protons at 1 GeV \protect\cite{Kaufman80}. Lower part: Charge distribution widths as a function of residue mass. 
The results of the same evaluations as in the upper part are presented (with unchanged notation),
except for protons at 1 GeV and at 3/6 GeV, for which the width parameter is equal to that obtained for
11.5/300 GeV \protect\cite{Kaufman80}.}
\label{aust_ach}
\end{figure}

The information on the reaction mechanism, obtained from the investigation of mass yield distributions, may
be enriched by the examination of other features of the $Y(A,Z)$ yield topography. Two such properties 
were compared for results of gold fragmentation by antiprotons and other projectiles: the course of
the fitted $Z_{p}$ path and the charge dispersion width. Figure \ref{aust_ach} illustrates such 
an inspection for some cases quoted in Table \ref{projectiles}. To bring differences into prominence, 
we have recalculated $Z_{p}(A)$ to its distance $\delta _{stab}$ from the line modeling the beta 
stability valley $Z_{\beta}$ (defined as in the caption to Fig. \ref{aust_map}). Such a presentation was
earlier applied to study the distribution of products of gold projectile fragmentation on C and Al
targets \cite{Souliotis98}. To compare with our results we present $\delta _{stab}$ obtained for 
the inverse kinematic Au + p reaction at 800 MeV \cite{Rejmund01,Benlliure01} and for energetic protons
\cite{Kaufman76,Kaufman80}. Besides this, we have also plotted curves derived for the Au target from a general 
formulae describing the $Z_{p}(A)$ path for products of various medium and heavy target fragmentations 
induced with protons at the fragmentation limit \cite{Summerer90} and with stopped antiprotons \cite{Egidy91}. 
The lower part of Fig. \ref{aust_ach} shows the dependence of the charge distribution width on the 
product mass.

As can be seen from Fig. \ref{aust_ach}, the relative course of $Z_{p}$ and $Z_{\beta}$ for stopped antiprotons 
changes now much more dramatically than in Fig. \ref{aust_map}. The $\delta _{stab}$ calculated for more general 
$Y(A,Z)$ models \cite{Summerer90,Egidy91} is smooth due to the broad $A$ range fitting. $Z_{P}$ fitted 
for 11.5 and 300 GeV protons \cite{Kaufman76} lies very close to this curves in the evaporation residues region.
Surprisingly, $Z_{p}$ derived for the inverse reaction at smaller energy extends farther towards the 
neutron-deficient nuclei for heavy products. In the fission region the situation is reversed except for the lightest
products. The curve plotted for 3 and 6 GeV protons lies closer to the valley of stability, for 1 GeV protons only 
the fission region is represented as there are no fit parameters given in Ref. \cite{Kaufman80}. 

When $Y(A,Z)$ 
models are applied to shorter mass ranges a rather non-continuous behavior with segments of rapidly changing 
position and orientation is observed. This happens both for protons \cite{Sihver92} and heavy ions 
\cite{Kudo84,Loveland90a,Loveland90b} investigated with $\gamma$-ray spectroscopy technique and for 
heavy ion reactions on Au studied with the inverse kinematic technique \cite{Souliotis98}. Obviously, 
this situation cannot be ascribed merely to the uncertainties of the experimental data, even in the worst cases. 
In our case, the experimental data distribution in the $N-Z$ plane shown in Fig. \ref{aust_map} strongly
favors the segmentation of the evaporation region in fitting.

Generally, lower energy reactions lead to $Z_{p}$ running closer the valley of stability, with passages 
to the neutron-rich side for fission fragments. Antiproton data show a peculiar tendency: although 
$Z_{p}$ lies quite away from the valley of stability for evaporation residues, it does not reach such
a neutron-deficient region as the energetic protons. Such a behavior may be partially explained by the 
influence of shell effects observed in antiproton distribution for $N/Z$ crossings at 106/76, 82/64 
and 50/40. The $Y(A,Z)$ yield reaches local maxima in these regions, the most probable $Z$ goes towards 
the more neutron-deficient nuclei and the charge dispersion becomes broader, as can be seen from Fig.
\ref{aust_ach}. 

The width of the $Z$ distribution, $\sigma$, was found to decrease smoothly with decreasing $A$ when a generalized 
approach is used for protons \cite{Summerer90} or antiprotons \cite{Egidy91}. On the contrary, results
of fitting within shorter mass regions are again inconsistent with compilations using broad mass regions, 
with quite small widths for evaporation residues and with a large scatter of $\sigma$ for fission fragments
\cite{Loveland90a,Loveland90b,Sihver92}. For antiprotons stopped in Au the charge width is rather small 
in the evaporation region, especially in comparison with 800 MeV/nucleon Au on H data. On the other hand, 
products of fission induced by stopped antiprotons are distributed quite broadly, similarly to the proton reaction 
products. 

From a methodological point of view, results on $Z_{p}(A)$ and $\sigma (A)$ fitted in different ways are 
not consistent, even for protons at similar energies. The $Y(A,Z)$ modeling applied for wide mass regions may be
reasonable for limited experimental data and for generalization purposes, however, this approach washes out 
any possible feature of more discrete nature. Hence, the division of input data to some $A$ subregions 
should work better in detailed studies, especially for lower excitation reactions.

\subsection{Antiprotons stopped in various targets}
\label{sec:targets}

\begin{figure}[t]
\begin{center}
\includegraphics[width=\columnwidth]{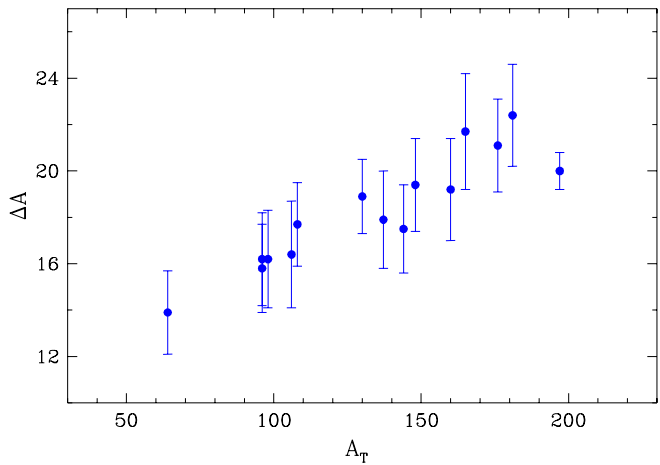}
\end{center}
\caption{Average mass removed from different targets after stopped antiproton absorption. Results from
earlier experiments studying mass yield curves 
\protect\cite{Moser89,Egidy90,Jastrzebski93,Szmalc92,Gulda93,Lubinski97},
are recalculated after yield correction for the heaviest products.}
\label{aust_dea}
\end{figure}

There were many other targets irradiated with low energy antiprotons from LEAR 
\cite{Moser89,Egidy90,Jastrzebski93,Szmalc92,Gulda93,Lubinski97}. A review of some results of these experiments 
will allow us to look closer at antiproton induced reactions. Mass-charge yield models were fitted only for a part 
of these targets, the parameters of the mass yield distributions for the rest were evaluated on the basis of summed 
direct experimental $Y(A)$ yields. However, either for the former or the latter results, the yields for the 
heaviest nuclei, close to the target, are underestimated since a significant part of the total Y(A) is hidden 
in non-detectable isotopes. To take this effect into account, we recalculated $\Delta A$ values obtained 
for other targets in the way as it was done for Au (see Sec.\ \ref{sec:projectiles}). The results are
listed in Table \ref{targets}. The removed mass increases with increasing target mass, as illustrated in 
Fig. \ref{aust_dea}. Such behavior is consistent with the simple geometrical picture of an excitation energy 
proportional to the number of participating nucleons \cite{Polster95}, hence to the volume of the nucleus bombarded 
with annihilation mesons.

\begin{table}[h]
\caption{Characteristics of mass yield distribution after antiproton stopping in different targets.}
\label{targets}
\begin{ruledtabular}
\begin{tabular}{ccc}
Target         & $\Delta A$    & Ref. \\
\colrule
$^{\rm nat}$Cu     & $13.9\pm 1.8$ & \phantom{}\cite{Jastrzebski93} \\
$^{96}$Ru      & $15.8\pm 1.9$ & \phantom{}\cite{Lubinski97} \\
$^{96}$Zr      & $16.2\pm 2.0$ & \phantom{}\cite{Lubinski97} \\
$^{98}$Mo      & $16.2\pm 2.1$ & \phantom{}\cite{Moser89} \\
$^{106}$Cd     & $16.4\pm 2.3$ & \phantom{}\cite{Lubinski97} \\
$^{\rm nat}$Ag     & $17.7\pm 1.8$ & \phantom{}\cite{Szmalc92} \\
$^{130}$Te     & $18.9\pm 1.6$ & \phantom{}\cite{Lubinski97} \\
$^{\rm nat}$Ba     & $17.9\pm 2.1$ & \phantom{}\cite{Egidy90} \\
$^{144}$Sm     & $17.5\pm 1.9$ & \phantom{}\cite{Lubinski97} \\
$^{148}$Nd     & $19.4\pm 2.0$ & \phantom{}\cite{Lubinski97} \\
$^{160}$Gd     & $19.2\pm 2.2$ & \phantom{}\cite{Lubinski97} \\
$^{165}$Ho     & $21.7\pm 2.5$ & \phantom{}\cite{Moser89} \\
$^{176}$Yb     & $21.1\pm 2.0$ & \phantom{}\cite{Lubinski97} \\
$^{\rm nat}$Ta     & $22.4\pm 2.2$ & \phantom{}\cite{Gulda93} \\
$^{\rm nat}$Au     & $20.0\pm 0.8$ & this work \\
\end{tabular}
\end{ruledtabular}
\end{table}

Using the $\Delta A$ value obtained for the Au target we may estimate the mean thermal excitation energy
of the decaying system. The compilation of the measured particle emission \cite{Markiel88,Hofmann90,Polster95} 
gives 5.4 
nucleons ejected in the cascade+preequilibrium stages through n, p, d, t, $^{3}$He and $^{4}$He ejectiles.
Hence we have on the average 14.6 evaporated nucleons and assuming 8 MeV separation energy and 3 MeV kinetic
energy per nucleon \cite{Polster95} leads to ($161\pm 23$) MeV stored in the thermalized system. Such a result
compares nicely with the value of ($183\pm 21$) MeV derived from the measurements of the spectra of neutrons 
and of light charged particles \cite{Polster95}.

\subsection{Odd-even effects}
\label{sec:shell}

Data on odd-even and shell effects observed in the $Y(A,Z)$ yield distribution are rather rarely 
discussed in the yield modeling context \cite{Rudstam66,Silberberg73,Egidy91}. Their influence 
on the yields is difficult to observe if the experimental data set is limited and errors are of the order 
of the possible
odd-even correction. The conclusion of Rudstam \cite{Rudstam66}, looking for a general formula predicting 
cross sections for p and $\alpha$ induced reactions, was that there is no need to introduce such
a correction as the experiment/model yield ratios for various N and Z combinations do not show
any clear correlation with the nucleon number evenness. Later, Silberberg and Tsao \cite{Silberberg73}
found a moderate effect, modeled with factors equal to 1.25, 0.9, 1.0 and 0.85 for even-even, odd-N, 
odd-Z and odd-odd (N,Z) pairs, respectively. 

Since off-line nuclear spectroscopy was applied to study the $Y(A,Z)$ distribution  
the odd-even effect was observed in reactions induced with stopped antiprotons \cite{Moser89,Egidy90}. 
Results for lighter targets ($^{92,95,98}$Mo, $^{\rm nat}$Ba) are consistent, with an 18-26 \% correction for 
odd-Z nuclei and a 32-34 \%  correction for odd-N nuclei (and the sum of these values in the odd-odd case). 
Corresponding values fitted for $^{165}$Ho \cite{Moser89} are not so evident, the yield of odd-N nuclei 
is strongly reduced (by $\approx$ 66\%) whereas there is no need to correct the odd-Z results. 
Because these fits were made simultaneously for the whole heavy residue region, no dependence on the 
emitted number of nucleons (hence excitation) was studied, also no indication for any shell effects 
was reported.

\begin{figure}[t]
\begin{center}
\includegraphics[width=\columnwidth]{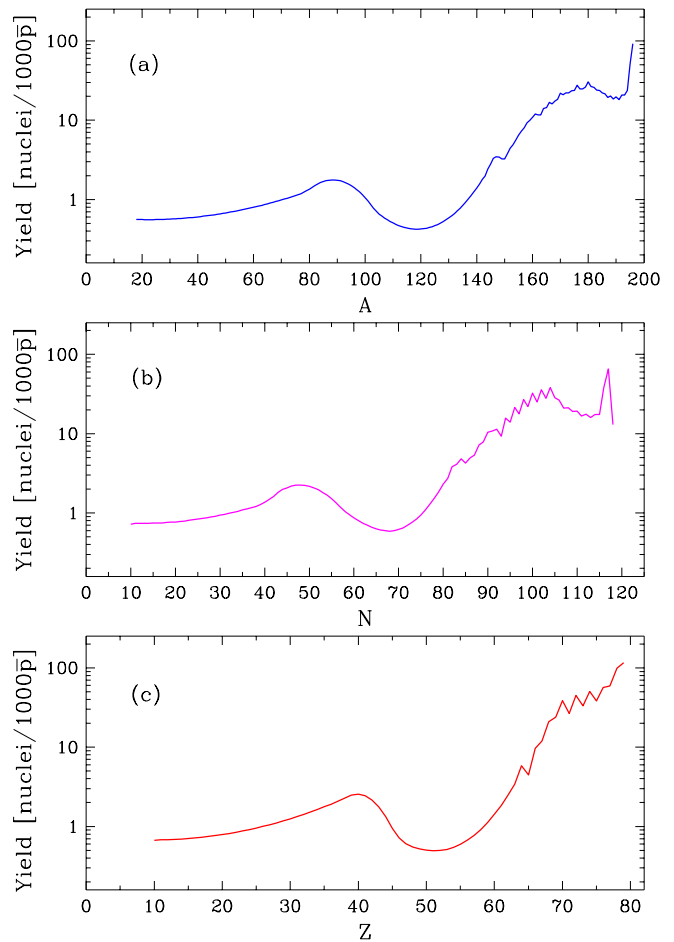}
\end{center}
\caption{Yield of residues after gold fragmentation by stopped antiprotons: (a) as a function of the atomic mass A,
(b) as a function of the neutron number N, (c) as a function of the atomic number Z.}
\label{aust_dis}
\end{figure}

Using the heavy and fissionable gold nuclei to absorb antiprotons we have the opportunity to investigate 
the odd-even and shell effects in a wide evaporation and fission products mass range. In Fig. \ref{aust_dis}
we present three fitted yield distributions, as function of mass, charge and neutron number of the
products. Corrections fitted for odd-N and odd-Z yield were shown in Table \ref{parameters} with parameters 
$a_{14}$ and $a_{15}$, 
respectively. They seem to be largest for the medium $\Delta A$ region ($161 < A < 182$), since between
the heavier residues only the odd-N Pt isotopes exhibit a clear yield reduction of about 30\%. A small 
odd-N effect ($\leq$ 10\%) seems also to appear for the other heavy even-Z products. Lighter evaporation
residues are produced more uniformly over changing nucleon numbers evenness, although for products close to the
closed $N=82$ shell ($143 < A < 149$) the even-N isotopes are strongly favored. We have not been able
to study odd-even effects for the lightest evaporation and all fission products because of their small
cross sections with large relative uncertainties and since the experimental data are there much less numerous. 
It should be stated that the correction for odd-odd nuclei used here slightly differs from those applied before
for stopped antiprotons \cite{Moser89,Egidy90} as we use the multiplicative form (Eq.\ (\ref{charge}))
instead of a correction factor equal to $1-P_{n}-P_{p}$, where $P_{n}$ and $P_{p}$ are parameters
fitted for odd-N and odd-Z nuclei, respectively. Since we observe that the odd-even effect is
stronger when both $N$ and $Z$ are odd or even, the multiplication is more adequate.

The use of odd-even corrections for stopped antiproton reactions strongly improves the fit,
with $\chi ^{2}$ reduction by more than one order of magnitude. Thus it cannot be treated as a trivial
improvement via adding another parameter. Moreover, since corrections obtained for different mass 
regions are consistent, therefore, taking into account previous results obtained for lighter targets 
\cite{Moser89,Egidy90} the inclusion of such a component is unavoidable in correct modeling of
the data coming from experiments with stopped antiprotons. Its strength, more pronounced than in the 
corresponding energetic proton data at 800 MeV \cite{Rejmund01,Benlliure01}, is one of the most distinct 
features of the antiproton absorption induced reactions.

\section{CONCLUSIONS}
\label{sec:conclusions}

The independent and/or cumulative yields for 114 isotopes produced after absorption of stopped
antiprotons in gold were measured by using the off-line $\gamma$-ray spectroscopy technique. On this
basis, with the help of a phenomenological model, the whole yield distribution was extracted for residues
ranging from the target mass minus one down to the light fission products with mass $\geq$ 20. The fission
probability was estimated to be ($3.8\pm 0.5$)\%, in agreement with the results of measurements 
using on-line techniques.

An average thermal excitation energy, gained by the Au nucleus after $\rm \overline{p}$ annihilation, was shown 
to be quite similar to that of 1 GeV protons, although the fission probability for such protons is almost 
twice as large. Moreover, the inspection of the yield distribution over the $A$-$Z$ plane indicates a fairly 
peculiar character of the reaction induced with low energy antiprotons. The most probable $Z$ course is quite 
different, lying closer to the stability line and exhibiting a more complex shape. Furthermore, the charge 
dispersion 
over $Z$ does not compare with that observed for 0.8 GeV protons, being almost twice as narrow.

The average mass removal observed for various targets reacting with stopped antiprotons rises linearly with 
increasing target mass. This behavior is consistent with in-beam studies of the light particle emission. 
$\Delta A$ derived from mass yield data helps to complete such measurements, unable to detect 
charged particles of the lowest energy.

A clear odd-even and some shell effects distinguish evidently the reaction with stopped antiproton from those
induced with energetic protons. For the first time the dependence of this phenomenon on the residue mass
was studied. The strength of such effects seems to diminish with the excitation energy, although for 
long evaporation chains and fission products it may be unobserved due to scarce and uncertain data.

\acknowledgments

This work was supported by the Polish State Committee for Scientific Research and by the German 
Bundesministerium f$\rm \ddot{u}$r Forschung und Technologie, Bonn.

\end{document}